\documentclass[12pt]{article}
\usepackage[cp1251]{inputenc}
\usepackage[english
]{babel}

\begin{document}

\title{Diffusion Processes in Phase Spaces and Quantum Mechanics}
\author{Evgeny Beniaminov\\
e-mail: ebeniamin@yandex.ru\\
web site: http://beniaminov.rsuh.ru}
\date{}

\maketitle
\begin{abstract}
A diffusion process for charge distributions in a phase space is examined. The corresponding charge moves in a force field 
and under an action of a random field.  There are the diffusion motions for coordinates and for momenta. 
In our model, an inner state of the charge is defined by a complex vector. The vector rotates with a great constant angular velocity 
with respect to the proper time of the charge.  A state of the diffusion process is a (complex-valued) wave function  on the phase space. 
 As in quantum mechanics, we assume that, for the wave functions, the superposition principle holds. 
The diffusion process averages out vectors of inner states from different points of the phase space. 
A differential equation for this diffusion process is founded and examined. We demonstrate that the motion (described by this process) 
decomposes into a fast motion and a slow motion. The fast motion reduces an arbitrary wave function to a function from a subspace 
whose elements are parameterized by complex-valued functions of coordinates. The slow motion occurs in this subspace 
and it is described by the Schr\"odinger equation.  The parameters of the suggested model are estimated. 
The duration of the fast motion is of order $10^{-11}$ s.   
\end{abstract}
\section*{Introduction}
In \cite{maslov1, maslov2}, Maslov studied the problem of describing the motion of a charge distribution on a phase space under
 the action of the a random field on the charges. He assumed that the distribution depends on a hidden parameter, which varies 
under motion. In \cite{maslov1}, it was proved under certain assumptions (Theorem~3.2) that, if the initial charge distribution, 
which depends on coordinates and momenta, belongs to a certain subspace parameterized by complex-valued functions 
of coordinates, then the distribution does not leave this subspace, and the dynamics of such a system is described by the 
corresponding Schr\"odinger equation. The nature of the hidden parameter is inessential, but the dependence of the initial 
distribution on this parameter must be periodic with period of order $\hbar$, where $\hbar$ is a small parameter (the Planck constant).

Following \cite{maslov1}, we consider the diffusion of charge distribution, which depends on coordinates, momenta, and an angular 
parameter $\theta /\hbar$. We assume that the dynamics of the system with respect to coordinates and momenta is determined 
by a Hamiltonian and a heat diffusion process.

The initial state can be any complex-valued function on the phase space (it is not required to belong to a certain function subspace, 
as in Theorem~3.2 from \cite{maslov1}). We show that, in the model under consideration, fast and slow motions can be distinguished. 
Under a fast motion, the systems starts in an arbitrary state and transfers into a state belonging to a subspace whose elements 
are parameterized by complex-valued functions of coordinates. A slow motion, as in Maslov's theorem, occurs in this subspace, 
and such a motion is described by the Schr\"odinger equation.

In this paper, we also estimate the parameters of the suggested model.

\section {A description of the model}
We consider a mathematical model of a process whose state at each moment of time is determined by a wave function, that is, 
a complex-valued function $\varphi (x, p)$, where $(x,p) \in R^{2n}$ and $n$ is the dimension of the configuration space. 
We assume that the change of states with time $t$ in this procedure is determined by a certain (described below) Feynman 
path integral in the phase space.

Consider the diffusion process on the phase space $(x, p) \in R^{2n}$ determined by a Fokker--Planck equation of the form
\begin{eqnarray}\label{eq_diffuz}
{ \frac{\partial f}{\partial{t}}=} \sum_{k=1}^n \biggl (\frac{\partial H}{\partial x_k} \frac{\partial f}{\partial p_k}-
\frac{\partial H}{\partial p_k} \frac{\partial f}{\partial x_k}
+a^2 \frac{\partial^2 f}{\partial x_k^2}+ b^2 \frac{\partial^2 f}{\partial p_k^2}\biggr),
\end{eqnarray}
where $f(x, p,t)$ is the probability density on the phase space at time $t$, 
$H(x, p)$ is the Hamiltonian function, and $a^2$ and $b^2$ are the coefficients of the diffusions with respect 
to the coordinates and momenta, respectively.

Diffusion process~(\ref{eq_diffuz}) determines in a standard way a probability measure 
${\cal P}[x^s, p^s]{\cal D} x {\cal D} p$ on the set of trajectories $\{ x^s, p^s\}$ going from the point 
$(x^0, p^0)$ at the initial moment $s=0$ and arriving at points $(x, p)$ at time $s=t$.

It is assumed that, if the wave function equals $\varphi^0 (x, p)$ at the initial moment $t=0$,
then its value $\varphi (x, p, t)$ at time $t$ is determined by the following integral, which is an analogue of the 
Feynman path integral \cite{feinman}:
\begin{eqnarray}\label{intFeinman}
 {\varphi (x, p, t) =}
e^{{abnt}/{\hbar}}\!\!\!
\int\limits_{R^{2n}}\!\!\!\varphi^0 (x^0, p^0)\!\!\!\!\! \int\limits_{(x^0,\ p^0)}^{(x,\ p)}\!\!\!\!\! F [x^s, p^s] 
{\cal P }[x^s, p^s] 
{\cal D} x {\cal D} p\ dx^0 dp^0,
\end{eqnarray}
\begin{equation}\label{intFeinmanFunk}
\mbox{where }\ \ \ \ \ F [x^s, p^s] = \exp\biggl\{-\frac {i}{\hbar} 
\int\limits_0^t \bigl( H(x^s, p^s) 
- \dot {x}^s p^s
\bigr)ds \biggr\},
\end{equation}
and the integral is with respect to the probability measure 
${\cal P }[x^s, p^s] {\cal D} x {\cal D} p$ on the set of trajectories $\{x^s, p^s\}$ defined above. 
The normalizing multiplier $\exp ({abn}t/{\hbar})$ of the integral is introduced for technical reasons.

We emphasize that integral~(\ref{intFeinman}) under consideration differs from the Feynman integral \cite{feinman} 
considered in quantum mechanics. In the case under consideration, the wave function depends on the coordinates 
and momenta, while in quantum mechanics, it depends only on the coordinates or only on the momenta. 
The quantum-mechanical Feynman integral is with respect to a "measure" defined by Feynman on the set of all trajectories 
in the configuration space. In the case under consideration, the integral is with respect to a probability measure on trajectories
 in the phase space. This probability measure is determined by the diffusion process~(\ref{eq_diffuz}).

As in quantum mechanics, we assume that, for wave functions, the superposition principle holds, and the probability density
 $\rho (x,p)$ on the phase space that corresponds to a wave function $\varphi(x, p)$ is defined by the standard formula
\begin{equation}\label{rho}
\rho (x,p)= \varphi^*(x, p) \varphi(x, p)=|\varphi(x, p)|^2. 
\end{equation}

Note that, like the Fokker--Planck equation, which is a continuous approximate model (for sufficiently large times) 
of a more exact discrete model of a diffusion process, the integral representation~(\ref{intFeinman}) can be regarded 
as a continuous approximation of a more exact discrete model, which is described by a complex Markov process. 
Such processes were studied by Maslov in\cite{maslov_mark}. 
In the case under consideration, the complex coefficients of the transition 
$u(x^{i+1}, p^{i+1}; x^i, p^i; \Delta t^i )$ from the point $(x^i, p^i)$ 
to the point $(x^{i+1}, p^{i+1})$ in time $\Delta t^i$ are given by
$$ 
u(x^{i+1}, p^{i+1}; x^i, p^i; \Delta t^i) = e^{Z_i} K(x^{i+1}, p^{i+1}; x^i, p^i; \Delta t^i ),
$$
$$
\mbox{where  }\ \ \ Z_i =  \frac{abn\Delta t^i}{\hbar}
- \frac{i}{\hbar}\biggl( H(x^i, p^i) \Delta t^i-(x^{i+1}-x^i)p^i \biggr), 
$$
and $K(x^{i+1}, p^{i+1}; x^i, p^i; \Delta t^i )$ is the probability of the transition from $(x^i, p^i)$ to $(x^{i+1}, p^{i+1})$ 
in time $\Delta t^i$ for a discrete diffusion process, which is described by the Fokker--Planck equation~(\ref{eq_diffuz}) 
in the limit as $\Delta t^i$ tends to zero.

\section{Main resalts}
For integral~(\ref{intFeinman}), a differential equation is constructed by a standard scheme~\cite{feinman}.

{\it{\bf Theorem 1.} The wave function determined by integral~(\ref{intFeinman}) satisfies the differential equation
\begin{equation}\label{eq_diff}
\frac{\partial\varphi}{\partial{t}}=\sum_{k=1}^{n}
\biggl(
\frac{\partial H}{\partial x_k} \frac{\partial\varphi}{\partial p_k}-
\frac{\partial H}{\partial p_k} \frac{\partial\varphi}{\partial x_k}
\biggr)
-\frac{i}{\hbar}
\biggl(H-\sum_{k=1}^{n}\frac{\partial H}{\partial p_k}p_k\biggr)\varphi
+\Delta_{a,b}{\varphi},
\end{equation}
\begin{equation}\label{delta}
\mbox{where }\ \ \ \ \ \Delta_{a,b}{\varphi}=
a^2\sum_{k=1}^{n}\biggl(\frac{\partial}{\partial{x_k}}-
     \frac{ip_k}{\hbar}\biggr)^{2}\varphi
+b^2\sum_{k=1}^{n}\frac{\partial^2 }{\partial{p^2_k}}\varphi
     +\frac {abn}{\hbar}{\varphi}.
\end{equation}}
Removing the last term from Eq.~(\ref{eq_diff}), we obtain a first-order partial differential equation. 
This part of Eq.~(\ref{eq_diff}) describes the deterministic component of the motion of complex vectors 
$\varphi(x, p,t)$. According to the equation, under this motion, the application point of each vector moves along the classical 
trajectory determined by the Hamiltonian $H(x, p)$, and the vector itself rotates with angular velocity 
$\omega'= \frac{1}{\hbar}\biggl(H-\sum_{k=1}^{n}\frac{\partial H}{\partial p_k}p_k\biggr)$ at each point of the trajectory.

Note that, if the configuration space is three-dimensional and $H=\\ c\sqrt {m^2c^2 +p^2}$, then 
$\omega' dt  =
\frac{mc^2}{\hbar}\frac{mc^2dt}{H}= \frac{mc^2}{\hbar} d\tau ,$
where $\tau$ is the proper time in the coordinate system related to a particle moving with momentum $p$. 
Thus, in this case, the vector, whose application point moves along a classical trajectory, rotates with constant angular velocity 
$\omega ={mc^2}/{\hbar}$ in the coordinate system related to this point.

Conversely, leaving only the last term of the form~(\ref{delta}) on the right-hand side of Eq.~(\ref{eq_diff}), we obtain the equation
\begin{equation}\label{eq_delta}
\frac{\partial\varphi}{\partial{t}}=
a^2\sum_{k=1}^{n}\biggl(\frac{\partial}{\partial{x_k}}-
     \frac{ip_k}{\hbar}\biggr)^{2}\varphi
+b^2\sum_{k=1}^{n}\frac{\partial^2 }{\partial{p^2_k}}\varphi
     +\frac {abn}{\hbar}{\varphi}.
\end{equation}
This equation describes the diffusion component of the motion of the vectors $\varphi(x, p,t)$ in the phase space. 
Under this motion, the application points of vectors move in accordance with the classical homogeneous diffusion process. 
Under small displacements from a point $(x, p)$ to a point $(x+dx, p+dp)$, each vector is parallel-translated, 
and its length at time $t$ is multiplied by $\exp (abnt/\hbar)$. 
The parallel translation of the vectors on the phase space is determined by the connection defined by 
$
L_{(dx,dp)}\varphi(x, p) - \varphi(x, p)\approx  -({i}/{\hbar})\varphi(x, p) p dq,
$
where $L_{(dx,dp)}\varphi(x, p)$ is the translation of the vector $\varphi(x, p)$ from the point $(x, p)$ 
along the infinitesimal vector $(dx, dp)$. In the special case where the configuration space is three-dimensional, such a connection 
on the phase space is caused by the synchronization of the moving clocks at the points of the phase space.
The right-hand side of ~(\ref{eq_delta}) is a self-adjoint operator. The eigenvalue problem for this operator reduces to 
the stationary Schr\"odinger equation for harmonic oscillations by applying the Fourier transform with respect to coordinates.
 This implies that the eigenvalues of the operator of Eq.~(\ref{eq_delta}) are nonpositive, and the following theorem is valid.

{\it {\bf Theorem 2}. Let $\varphi(x,p,0)$ be an arbitrary function whose Fourier transform with respect to $p$ tends to zero 
as $x \rightarrow  \infty $.
Then, the solution $\varphi(x,p,t)$ to the diffusion equation~(\ref{eq_delta}) tends to a stationary solution of the form
\begin{equation}\label{view_varphi}
\varphi(x,p)=\lim_{t \to \infty} \varphi(x,p,t)=
\frac{1}{(2\pi{\hbar})^{n/2}}
\!\int\limits_{R^n}\!\!\psi(y)\chi(x,y)
e^{-{{i  (y-x)p}/{\hbar}}}
dy,
\end{equation}
\begin{eqnarray}\label{psi}
 \mbox {where }\ \ \psi(y) =\frac{1}{(2\pi{\hbar})^{n/2}}
\int\limits_{R^{2n}}\!\!\varphi(x,p,0)
e^{{{i  (y-x)p}/{\hbar}}}
\chi(x,y)dpdx,
\end{eqnarray}
\begin{eqnarray}\label{chi}
 \mbox{and          }\ \ \ \ \ \ \ \ \ \ \ \ \ \ \ \ \ 
  \chi(x,y)=\left(\frac{b}{a\pi\hbar}\right)^{n/4}e^{-{{b}(x-y)^2}/{(2a\hbar)}},
\end{eqnarray}
exponentially in time (with exponent $-abt/{\hbar}$).
}

Note that $\chi^2(x,y)$ is the probability density of the normal distribution in $x$ with mathematical expectation 
$y$ and variance $a\hbar/(2b)$. If the value of $a\hbar/(2b)$ is small, then the function $\chi^2(x,y)$ is close to the delta-function 
of $x-y$.

The composition of~(\ref{psi}) and (\ref{view_varphi}) is a projector from the space of wave functions on the phase space onto 
some subspace. The elements of this subspace are parameterized by functions of the form $\psi(y),$ where $y\in R^n$, 
i~e., by wave functions on the configuration space.

Formulas~(\ref{view_varphi}) and (\ref{rho}) imply the following assertion.

{\it {\bf Theorem 3}. If $\psi(x)$ is a wave function on the configuration space, then the corresponding probability density 
on the phase space is determined by
\begin{equation}\label{rho_psi}
\rho (x,p)=
\frac{1}{(2\pi{\hbar})^{n}}
\int\limits_{R^{2n}}\!\!\psi(y)\psi^{*}(y')\chi(x,y)\chi(x,y')
e^{-{{i (y-y')p}/{\hbar}}}
dy dy',
\end{equation}
where $\chi(x,y)$ is defined by~(\ref{chi}).
}

In contrast to the quasi-distributions defined by Wigner in~\cite{vigner}, the density $\rho(x,p)$ defined on the phase space 
by~(\ref{rho_psi}) is always nonnegative. The algebra of observables determined by real functions on the phase space 
but averaged over probability distribution densities of the form~(\ref{rho_psi}) was studied in~\cite{ben}.

The probability distribution density $\rho(x)$ on the configuration space is expressed in terms of the density $\rho(x,p)$ 
by the formula $\rho(x)=\int_{R^3}\rho(x,p) dp.$ Using this expression and integrating~(\ref{rho_psi}) with respect to $p$, 
we obtain the following assertion.

{\it  {\bf Corollary 1}. If $\psi(x)$ is a wave function on the configuration space, then the corresponding density 
$\rho(x)$ on the configuration space is given by
\begin{equation}\label{rho_x}
\rho (x)=\int_{R^3}|\psi(y)|^2 \chi^2(x,y) dy,
\end{equation}
where $\chi(x,y)$ is defined by~(\ref{chi}). Thus, $\rho (x)$ is obtained from $|\psi(x)|^2$ by smoothing (convolving)
 with respect to the density of the normal distribution with variance $a\hbar/(2b)$, and the accuracy of the determination 
of the coordinate is bounded by $\sim \sqrt{a\hbar/(2b)}$.
}

As is known\cite{landau4}, in quantum electrodynamics, the minimal error of measuring the coordinates of an electron in a system 
at rest is estimated as $ \hbar/(mc),$ where $m$ is the mass of the electron and $c$ is the speed of light.

In statistical physics (see, e.g., \cite{isihara}, Chapter 7, Sections 4 and 9), under the assumption that diffusion is caused by thermal 
actions on electrons, the diffusion coefficients in the coordinates and momenta are expressed in terms of the temperature $T$ as
$a^2= kT/(m\gamma) \ \  \mbox{      and     }\ \ b^2=\gamma kTm,$
where $k$ is the Boltzmann constant, $m$ is the mass of the electron, and $\gamma$ is the friction coefficient of the medium per unit mass. 
This implies $a/b=(\gamma m)^{-1}$ and $ab=kT$. Thus, in this case, the quantity $a/b$, which is contained in~(\ref{chi}) 
and determines the variance of smoothing in Corollary~1, does not depend on the temperature. 
Moreover, the duration $t$ of the transition process, which is defined in Theorem~2, has the form 
$t \sim {\hbar}/{(ab)}={\hbar}/{(kT)}= T^{-1}\cdot 7.638\cdot 10^{-12}\ s.$

Taking into account this estimate, we regard the quantity ${\hbar}/{(ab)}$ in Eq.~(\ref{eq_diff}) as a small parameter 
and assume that the coordinates and momenta little change during the transition time under the classical motion determined 
by the Hamiltonian $H(x, p)$.

{\it {\bf Theorem 4}. The motion described by Eq.~(\ref{eq_diff}) asymptotically decomposes into a fast and a slow motion as 
${\hbar}/{(ab)} \rightarrow   0$. The fast motion reduces an arbitrary wave function $\varphi(x, p, 0)$ to the form~(\ref{view_varphi}) 
in time of order ${\hbar}/{(ab)}$. The wave functions of the form~(\ref{view_varphi}) constitute a linear subspace. 
The elements of this subspace are parameterized by wave functions $\psi(y)$ depending only on the coordinates $y\in R^n$.
 The slow motion, which starts from a nonzero wave function belonging to this subspace, occurs in this subspace 
and is parameterized by a wave function $\psi(y,t)$ depending on time. The function $\psi(y,t)$ satisfies the 
Schr\"odinger equation $ i\hbar {\partial \psi}/{\partial t} = \hat{H}\psi$, where
\begin{eqnarray}
\hat{H}\psi & = & \frac {1}{(2\pi \hbar)^n} \int \limits_{R^{3n}}  
\biggl (H(x, p)-\sum_{k=1}^{n}\biggl(\frac{ \partial H}{\partial x_k}
+\frac {i b}{a}\frac{ \partial H}{\partial p_k}\biggr)(x_k-y'_k)
\biggr)\times \nonumber\\
   &  & \times \chi(x, y) \chi (x, y')  e^{\frac{i}{\hbar}(y-y') p}  \psi(y',t) dy' dx dp, \nonumber  
\end{eqnarray}
and $\chi(x, y) $ is defined by~(\ref{chi}).
}

{\it{\bf  Theorem 5}. If $\frac {a\hbar}{b}$ is a small quantity and $H(x, p) = \frac{p^2}{2m}+ V(x),$ 
then, accurate to terms of order $a\hbar/b$, the operator $\hat H$ has the form
\begin{equation}\label{hatH}
\hat H \approx - \frac{\hbar^2}{2m}\biggl(\sum_{k=1}^{n}\frac{\partial^2 }{\partial{y^2_k}}\biggr)
+V(y)-\frac{a\hbar}{4b}\sum_{k=1}^{n}\frac{\partial^2 V}{\partial{y^2_k}} +\frac{3nb\hbar}{4ma}.
\end{equation}
}

The first two terms in~(\ref{hatH}) give the standard Hamiltonian operator; the last term is a constant, and it can be neglected.
The next-to-last term can be regarded as a perturbation of the Hamiltonian (because $a\hbar/b$ is small).

Assuming that the deviations in the spectrum of the hydrogen atom (the Lamb shift), which are observed in the Lamb--Retherford 
experiments~\cite{lamb}, are caused by the next-to-last term in~(\ref{hatH}), we can estimate the quantity  $a/b$.
Calculations by the standard method of perturbations, similar to those performed in \cite{welt}, give the estimate 
$a/b =3.41\cdot 10^4 s/g$. Thus, the standard deviation for the normal distribution $\chi^2$, 
over which smoothing in~(\ref{rho_x}) is performed, has the form $\sqrt{a\hbar/(2b)}=4.24 \cdot 10^{-12} cm$. 
This quantity is substantially smaller than the radius of the hydrogen atom and close to the Compton wave length 
$\hbar/(mc)=3.86\cdot 10^{-11} cm$ of the electron.

Thus, calculations show that the suggested model, which has the form of integral~(\ref{intFeinman}) or of differential 
equation~(\ref{eq_diff}), sufficiently adequately describes physical processes in standard cases for the standard Hamiltonian. 
This model can also be applied to calculate processes with nonstandard Hamiltonians or with Hamiltonians rapidly varying with time, 
which take place under sudden perturbations~\cite{dyihne} or when the potential varies periodically with frequency of order 
$ab/\hbar$; it would be interesting to compare the results with experimental data.

\end{document}